\documentclass[a4paper,12pt]{article}
\usepackage{psfrag}
\usepackage{graphicx}
\usepackage{graphics}
\usepackage{bm}
\usepackage{color}
\usepackage{amssymb}
\usepackage{amsmath}
\input{epsf.sty}

\usepackage{vmargin}            
\setmarginsrb{3cm}{3cm}{3cm}{2cm}{0cm}{0cm}{0cm}{1cm}

\title{Multi-field DBI inflation: introducing  bulk forms and revisiting the 
gravitational wave constraints}

\author{David Langlois$^{1,2}$\footnote{langlois@apc.univ-paris7.fr}, S\'ebastien Renaux-Petel$^1$\footnote{renaux@apc.univ-paris7.fr} and Dani\`ele A.~Steer$^{1}$\footnote{steer@apc.univ-paris7.fr} \\
{\small {}}\\
{\small ${}^1${\it APC (Astroparticules et Cosmologie),}}\\
{\small {\it
UMR 7164 (CNRS, Universit\'e Paris 7)}}\\
{\small {\it  10, rue Alice Domon et L\'eonie Duquet,
 75205 Paris Cedex 13, France}}\\
 {\small ${}^2${\it Institut d'Astrophysique de Paris (IAP),}}\\
{\small {\it 98bis Boulevard Arago, 75014 Paris, France;  }}\\}

\begin{document}

\date{\today}
\maketitle

\newcommand{\ba}{\begin{eqnarray}}
\newcommand{\ea}{\end{eqnarray}}

\newcommand{\nc}{\newcommand}
\newcommand{\nn}{\nonumber}
\nc{\bea}{\begin{eqnarray}}
\nc{\eea}{\end{eqnarray}}
\newcommand\be{\begin{equation}}
\newcommand\beq{\begin{equation}}
\newcommand\ee{\end{equation}}
\newcommand\eeq{\end{equation}}
\newcommand\bb{{\mathbf{b}}}
\newcommand\baa{{\mathbf{a}}}
\newcommand\bee{{\mathbf{e}}}
\newcommand\bff{{\mathbf{f}}}
\newcommand\bx{{\mathbf{x}}}
\newcommand\barf{{\bar{f}}}
\newcommand\barp{{\bar{p}}}
\newcommand\barphi{{\bar{\phi}}}
\newcommand\barR{{\bar{R}}}
\newcommand\si{\sigma}
\newcommand\al{\alpha}
\newcommand\ta{\tau}
\newcommand\Tr{{\rm Tr}}
\nc{\ga}{\gamma}
\nc{\tr}{{\rm{tr}}}
\nc{\tnu}{\bar{\nu}}
\nc{\tmu}{\bar{\mu}}
\nc{\tq}{\tilde{q}}
\def\ap{{\alpha^{\prime}}}
\newcommand{\la}{\lambda}
\nc{\x}{{\bf{x}}}
\newcommand{\gsim}{\raise.3ex\hbox{$>$\kern-.75em\lower1ex\hbox{$\sim$}}}
\newcommand{\lsim}{\raise.3ex\hbox{$<$\kern-.75em\lower1ex\hbox{$\sim$}}}
\nc{\gm}{\gamma }
\nc{\PIJ}{\tP_{,X_{IJ}}}
\nc{\hP}{\hat{P}}
\nc{\PX}{\tP_{,X}}
\nc{\half}{\frac{1}{2}}
\nc{\p}{\phi}
\newcommand{\dn}[2]{{\mathrm{d}^{{#1}}{{#2}}}}
\nc{\s}{\sigma}
\newcommand{\mat}[1]{\underline{\underline{#1}}}
\newcommand\bk{\boldsymbol{k}}
\nc\pp{\psi}
\nc\bfphi{{\bf \phi}}
\def\a{\alpha}
\def\b{\beta}
\def\H{{\cal F}}
\def\tP{{\tilde P}}
\def\tX{{\tilde X}}
\def\tG{{\tilde{g}}}
\def\e{e}
\def\A{{A_0}}
\def\Ai{A_i}
\def\M{A}
\def\B{ B}
\def\L{H}
\def\te{\tilde e}
\def\c{c}
\def\d{\Phi}
\def\pp{\beta}
\def\m{\mu}
\def\hg{\hat{\gamma}}

\def\Tdot#1{{{#1}^{\hbox{.}}}}
\def\Tddot#1{{{#1}^{\hbox{..}}}}

\def\fF{{\lambda}}
\def\hB{{\hat B}}
\def\hG{{\hat G}}

\def\C{{\cal B}}
\def\Bc{{\cal F}}

\def\S{{\bf S}}

\def\R{{\cal R}}

\def\KR{b}

\def\c{{\rm compact}}
\def\nc{{\rm non}$-${\rm compact}}
\def\pb{{\bar \phi}}
\def\Ni{N_{init}}
\def\Nf{N_{fin}}

\begin{abstract}
We study multi-field Dirac-Born-Infeld (DBI) inflation models, taking into account the NS-NS and R-R bulk fields present in generic flux compactifications.
We compute the second-order action, which governs the behaviour of linear cosmological perturbations, as well as
the  third-order action, which can be used to calculate non-Gaussianities in these models. 
Remarkably, for scalar-type perturbations, we show that the contributions  due to the various form fields  exactly cancel in both the second- and third-order actions.  Primordial perturbations and their non-Gaussianities are therefore unaffected by the presence of form fields and our previous results are unmodified.
We also study vector-type perturbations associated with the U(1) gauge field  confined on the D3-brane, and discuss 
whether their quantum fluctuations can be amplified.  
Finally, we revisit the gravitational wave constraints on DBI inflation and show that an ultra-violet DBI multi-field scenario is still compatible with data, in contrast with the single field case, provided there is a transfer from entropy into adiabatic perturbations.
\end{abstract}

\newpage

\section{Introduction}

Although inflation is today the main mechanism  with which to describe the very early universe, the nature of the field or fields responsible for inflation still remains an open question. In particular, inflation model building in the context of string theory has proved rather challenging (for recent reviews, see e.g.~\cite{McAllister:2007bg,Burgess:2007pz,Kallosh:2007ig,Cline:2006hu,HenryTye:2006uv}), one of the reasons being that 
inflation usually requires an extremely flat potential. An interesting way to bypass this problem\footnote{Recently, the slow down of scalar fields by particle production was also revived in \cite{Barnaby:2009mc,Green:2009ds}.}  is to resort 
to non-standard kinetic terms  e.g.~\cite{ArmendarizPicon:1999rj,Gibbons:2002md}, and these arise naturally in inflation associated 
with the motion of a D-brane in a higher-dimensional spacetime.  Indeed, the Dirac-Born-Infeld (DBI) action, which is a part of the full D-brane action, contains non-standard kinetic terms of a specific form.  In the context of warped geometries with moving D3-branes, 
they can lead to a phase of inflation, dubbed DBI-inflation \cite{st03,ast04,Chen:2004gc,Chen:2005ad}.

In DBI-inflation, the coordinates of the brane in the  higher-dimensional spacetime give rise to scalar fields from the effective four-dimensional point of view. 
Although initial works studied the motion of the brane along a {\it single direction} (namely the radial direction 
in the context of warped conical compactifications), it is interesting to consider the more general motion of the brane in all compact dimensions (namely the angular directions also).
  This effectively leads to {\it multiple field} inflationary scenarios.

DBI inflation, and more generally inflation with non standard kinetic terms, can lead to significant non-Gaussianity which could potentially be observable in the future measurements of the CMB by the Planck satellite.    
The cosmological perturbations generated in such scenarios have been investigated in single-field inflation 
 (see e.g. \cite{Garriga:1999vw,Chen:2006nt} for a general analysis), and in
multi-field inflation
 \cite{Langlois:2008mn,Langlois:2008qf,Langlois:2008wt,Arrojaetal}. 
In the latter case,  entropy modes can be generated during inflation, in addition to the usual adiabatic modes,  and these entropy modes can affect the final  curvature  perturbation if there is a transfer from entropy modes into adiabatic modes. As we also showed  in the context of multi-field DBI inflation, this transfer also affects the amplitude of non-Gaussianities, although the shape is the same as in the single-field case (assuming a small  sound speed).

In our previous works on multi-field DBI-inflation \cite{Langlois:2008qf,Langlois:2008wt}, we ignored  the presence of NS-NS and R-R form fields in the bulk, as well as the $U(1)$ gauge field confined on the brane:
these fields, however, are generically present in typical bulk solutions, for example the Klebanov-Strassler solution \cite{Klebanov:2000hb,Herzog:2001xk}, and they contribute to the D3-brane action through both the DBI and Wess-Zumino terms.
One of the goals of  this paper is thus to  investigate the consequences  of the bulk form fields on scalar-type cosmological perturbations, both at linear and non-linear order. We find that for the second-order and third-order actions involving  scalar 
perturbations, the terms arising from the coupling between the bulk forms and the brane position scalar fields are exactly compensated by the terms due to the fluctuations of the $U(1)$ gauge field confined on the brane. This implies that our  previous results for the primordial spectra and the non-Gaussianities remain unchanged in this more general context.
We also study the two vector degrees of freedom associated with the $U(1)$ gauge field on the brane and investigate in which case their quantum fluctuations can be amplified during inflation. 

Finally,   we consider the observational constraints on DBI-inflation models, based on a combination of their predictions for the scalar spectral index, non-Gaussianities and gravitational waves. By confronting an upper bound on the amount of gravitational waves \cite{Lyth:1996im} (typically negligible in the original models of DBI inflation \cite{Baumann:2006cd}) with a lower bound related to the deviation of the scalar spectrum from scale-invariance \cite{Lidsey:2007gq}, it has been argued that most models of {\it single-field}
ultra-violet DBI inflation are ruled out.  In the {\it multi-field} case, however, the second constraint can be relaxed when there is  a transfer from  entropy into adiabatic modes meaning that these models are not excluded.

The plan of the present paper is the following. In the next section we derive the effective multifield DBI action, paying particular attention to the various bulk form fields. In Sec.~\ref{sec:background} we focus on the homogeneous background evolution. The second-order action is calculated in Sec.~\ref{sec:secondorder}, and in Sec.~\ref{sec:linear-scalar} we focus on linear scalar perturbations, comparing our results to those of \cite{Langlois:2008wt}. In Sec.~\ref{sec:vector} we concentrate on the vector perturbations associated with the abelian gauge field confined on the brane.
Sec.~\ref{sec:NG} is devoted to the predictions for non-Gaussianity, based 
on the third-order action. In Sec.~\ref{sec:GW-constraints}, we revisit arguments which disfavour single field ultra-violet DBI inflation, and show how a multi-field scenario can overcome those difficulties. We draw our conclusions in Sec.~\ref{sec:conclusion}.

\section{The effective multi-field DBI action}
\label{sec:action}

In this section we derive the four-dimensional effective action associated with the motion of a 
probe D3-brane moving through a compact space with coordinates $y^K$ ($K=1,\ldots,6$). The $\nc$ space-time coordinates are denoted by $x^\mu$ ($\mu=0,1,2,3$).

Although we do not use 
any specific geometry and remain as general as possible, we have in mind warped flux compactifications
in type IIB string theory \cite{Giddings:2001yu} and consider a ten-dimensional metric of the general form 
\ba
ds^2 &=& h^{-1/2}(y^K)\,g_{\mu \nu}(x^\lambda) \, dx^\mu dx^\nu + h^{1/2}(y^K)\, \tG_{IJ}(y^K)\, dy^I dy^J 
\nonumber
\\
&\equiv& \gamma_{AB}\,  dY^A dY^B\,  .
\label{metric1}
\ea
Here $Y^A=\left\{x^\mu, y^I\right\}$, and the warp factor $h$ and metric $\tG_{IJ}$ depend only on the $\c$ coordinates $y^K$. In a general flux compactification, the dilaton $\d$ may be non trivial and all fluxes may be turned on: the R-R forms $F_{n+1}=d C_n$ for $n=0,2,4$ (and their duals), as well as the NS-NS flux $H_3=d B_2$. In order to maintain four-dimensional local Lorentz invariance, the three-fluxes $F_3$ and $H_3$ have only $\c$ components and the axion $C_0$ and dilaton $\d$ are allowed to vary only along the $\c$ manifold. 
As a consequence, in the following, we choose a gauge in which $C_2$ and $B_2$ have non-trivial components only along the $\c$ directions, whereas $C_4$ has components only along the $\nc$ spacetime dimensions. 
 Thus
\ba
 B_{IJ} \neq 0    &\qquad&  B_{\mu I}=B_{\mu \nu}=0 \, ,
\label{Bsym}
\\
 C_{IJ} \neq 0  &\qquad&  C_{\mu I}=C_{\mu \nu}=0 
\, ,
\nonumber
\\
C_{\mu \nu \rho \sigma} \neq 0 & & {\rm with \; all \; other \; components \; vanishing.}
\ea
 
 The action for a single D3-brane, with tension $T_3$,  in this background is
\be
S_{\rm brane} = S_{\rm DBI} + S_{\rm WZ}
\ee
where the Dirac-Born-Infeld (DBI) and Wess-Zumino (WZ) actions are given by \cite{Leigh:1989jq,Bachas:1998rg}
\ba
\label{L}
S_{\rm DBI} &=& -T_3\int {\rm d}^4 x\,  e^{-\d}\sqrt{-\det{ \left(\hat{\gamma}_{\mu \nu}+\hat B_{\mu \nu} +2 \pi \alpha' F_{\mu \nu}\right) }}
\\
S_{\rm WZ}&=& - T_3\int_{\rm brane} \sum_{n=0,2,4} \left. \hat{C}_n\wedge e^{\left({\hat B}_2 +2 \pi \alpha' F_2 \right) } \right|_{4{\rm -form}}\, .
\label{WZ}
\ea
Here and in the following, a hat denotes a pull-back onto the brane so that $\hat{\gamma}_{\mu \nu}$, for example, is the induced metric on the brane. In the Wess-Zumino term, one keeps only the $4$-forms resulting from the wedge product, and $F_2$ is the field strength of the worldvolume $U(1)$ gauge field, i.e. $F_{\mu \nu} \equiv \partial_{\mu } A_{\nu}-\partial_{\nu } A_{\mu}$. Note that it enters the brane's action only through the combination
\beq
\Bc_{\mu \nu}=\hat{B}_{\mu \nu}+2 \pi \alpha'F_{\mu \nu}\,
\label{combination-new}
\eeq
as is required from gauge invariance under
the NS gauge transformation $B_2 \to B_2 +d \Sigma_1$, $2 \pi \alpha' F_2 \to 2 \pi \alpha'F_2 -\widehat {d \Sigma_1}$ \cite{Polchinski}. Thus only $\Bc_{\mu \nu}$ as a whole is the physical, i.e.~gauge-invariant, field strength on the brane. This implies, in particular, that one cannot take into account the perturbations of $\hat{B}_{\mu \nu}$ without doing so for $F_{\mu \nu}$: this point will have important consequences below.
Finally, the  brane embedding is defined by the functions  
\be
Y_{\rm (b)}^A(x^\mu)= (x^\mu, \varphi^I(x^{\mu}))
\label{brane-embed}
\ee
where we have chosen the brane spacetime coordinates $x^\mu$ to coincide with the first four bulk coordinates. 

It will be useful for the following discussion to consider the DBI and WZ actions separately.

\subsection{DBI action}

Using (\ref{metric1}) and (\ref{Bsym}), the metric and two-form induced on the brane are given by
\ba
\hat{\gamma}_{\mu \nu} &=&  \gamma_{AB}\, \partial_\mu Y_{\rm (b)}^A \partial_\nu Y_{\rm (b)}^B  = h^{-1/2} \left( g_{\mu \nu}  + h \tG_{IJ} \partial_\mu \varphi^I \partial_\nu \varphi^J \right)
\label{met}
\\
\hat B_{\mu \nu} &=&  B_{AB}\,  \partial_\mu Y_{\rm (b)}^A \partial_\nu Y_{\rm (b)}^B\, = 
 \B_{IJ}\,  \partial_\mu \varphi^I \partial_\nu \varphi^J \,.
\label{KBzero}\,
\ea
Now let us use the following rescalings
\beq
\phi^I = \sqrt{T_3}\, \varphi^I \; , \quad  G_{IJ}=e^{- \d}\, \tG_{IJ} \, \quad \KR_{IJ}=\frac{h^{1/2}}{T_3}\, B_{IJ}\, .
\label{redef}
\eeq
Then on defining the functions
\beq
\label{f}
f(\phi^K)\equiv e^{\d} \frac{h}{T_3}\,, \qquad  \fF(\phi^K)=2\pi \alpha' h^{1/2}
\eeq
which depend only on the scalar fields,
we can rewrite the DBI action Eq.~(\ref{L}) as
\beq
 S_{\rm DBI} =\int d^4x \sqrt{-g} \left(-\frac{1}{f}\sqrt{{\cal D}}\right),
\eeq
with the determinant  
\beq
{\cal D}\equiv 
\det(\delta^{\mu}_{\nu}+f\, G_{IJ} \partial^{\mu} \phi^I \partial_{\nu} \phi^J+b_{IJ} \partial^{\mu} \phi^I \partial_{\nu} \phi^J +\fF F^{\mu}_{\ \nu} )\, ,
\label{D}
\eeq
where the greek indices are raised and lowered with the `spacetime' metric $g_{\mu \nu}$.
We can rewrite this determinant in the form 
\beq
{\cal D}=  \det({\bf I}+{\bf S}+\bf{\C}),  
\label{Dmatrix}
\ee
where ${\bf I}$ is the four-dimensional identity matrix, and from Eq.~(\ref{D}) the matrices ${\bf S}$ 
and ${\bf \C}$ are defined through their components
\beq
S^{\mu}_{\, \nu}=f {G}_{IJ} \partial^{\mu} \p^I \partial_{\nu} \p^J \qquad (S_{\mu\nu} = S_{\nu \mu})
\label{def-S}
\eeq
and
\beq
\C^{\mu}_{\ \nu}= 
b_{IJ} \partial^{\mu} \p^I \partial_{\nu} \p^J+ \fF F^{\mu}_{\; \nu} \,=\frac{\fF}{2\pi \alpha'} g^{\mu \lambda}\Bc_{\lambda \nu} \qquad (\C_{\mu\nu}= - \C_{\nu \mu})
\label{def-A}
\eeq 
 where $\Bc_{\mu \nu}$ was defined in Eq.~(\ref{combination-new}).

Computing the determinant in Eq.~(\ref{Dmatrix}) yields  
\begin{eqnarray}
{\cal D}&=& {\cal D}_{S}
-\half \Tr(\C^2) \left(1+\Tr \S \right)  +\Tr(\S \C^2) \left(1 + \Tr \S\right)  -\Tr(\S^2 \C^2)
\nonumber\\
 &&~~{}
-\frac{1}{4}\Tr(\C^2)[(\Tr \S)^2-\Tr(\S^2)]-\half \Tr(\S\C \S\C)
 \nonumber\\
 &&~~{}+\frac{1}{8}\left[ \left(\Tr(\C^2)\right)^2-2\Tr(\C^4) \right]
\label{D-explicit}
\end{eqnarray}
where
\be
{\cal D}_{S} \equiv 1+\Tr \S+\half [(\Tr \S)^2-\Tr(\S^2)]+ S_{\alpha}^{[\alpha}  S_{\beta}^{\beta} S_{\gamma}^{\gamma ]} + S_{\alpha}^{[\alpha}  S_{\beta}^{\beta} S_{\gamma}^{\gamma } S_{\delta}^{\delta ] } \, .
\label{D_S}
\ee
When $\S$ vanishes,  ${\cal D}$ reduces to the determinant of standard Born-Infeld theory \cite{Born:1934gh}. On the other hand, when
 the brane and bulk form fields are ignored, 
${\cal D}$ reduces to ${\cal D}_{S}$, which depends only on the scalar fields. As we showed in \cite{Langlois:2008qf,Langlois:2008wt}, it can be written in the form
\bea
{\cal  D}_S= 1-2f G_{IJ}X^{IJ} + 4f^2 X^{[I}_IX_J^{J]} 
-8f^3 X^{[I}_IX_J^{J} X_K^{K]}+16f^4 X^{[I}_IX_J^{J} X_K^{K}X_L^{L]},
\eea
where we have defined 
\beq
X^{IJ}\equiv -\frac{1}{2} \partial^{\mu} \p^I \partial_{\mu} \p^J , \quad X_I^J=G_{IK}X^{KJ},
\eeq
and where the brackets denote antisymmetrisation of the field indices.  Above and in the following, field indices are raised and lowered with the field space metric $G_{IJ}$ defined in (\ref{redef}).

\subsection{WZ action}

Let us now turn to the  Wess-Zumino part. The explicit expression given in Eq.~(\ref{WZ}) is 
\bea
S_{\rm WZ}&=&-T_3 \left[\int_{\rm brane} \hat{C}_4+ \int_{\rm brane} \hat{C}_2\wedge \left(\hat{B}_2 +2 \pi \alpha' F_2 \right) \right.
 \cr &&
 + \left. \frac12 \int_{\rm brane} C_0\left(\hat{B}_2\wedge \hat{B}_2 +4 \pi \alpha' \hat{B}_2\wedge F_2+(2\pi\alpha')^2F_2\wedge F_2\right) \right]
 \nonumber
 \\
&\equiv&  S_{\rm WZ}^{[4]} + S_{\rm WZ}^{[2]} \, ,
\label{WZexpanded}
\eea
where we have separated out the part ($S_{\rm WZ}^{[4]}$) coming from the 4-form $\hat{C}_4$, and the remainder  ($S_{\rm WZ}^{[2]}$) containing the 2-forms $\hat C_2, \hat B_2$ and $F_2$.
The 4-form $\hat{C}_4$ is given by 
\beq
\hat{C}_4={\cal V}\,  \epsilon_4,
\eeq
where $\epsilon_4$ is the fully antisymmetric tensor associated with the four-dimensional metric $g_{\mu\nu}$ (so 
that $\epsilon_{0123}=\sqrt{-g}$) and the coefficient ${\cal V}$ depends only on the compact coordinates.
Therefore the first term on the right hand side of Eq.~(\ref{WZexpanded}) yields, in the effective four-dimensional action,
a potential term depending on the scalar fields $\phi^I$ characterizing the brane position in the $\c$ space:
\be
S_{\rm WZ}^{[4]} = -T_3 \int_{\rm brane} \hat{C}_4 = - \int {\rm d}^4 x\, \sqrt{-g} \, T_3  {\cal V} (\phi^I)\, .
\label{pot}
\ee
There are five terms in $S_{\rm WZ}^{[2]}$. The first, involving $\hat{C}_2$, is proportional to
\ba
\int_{\rm brane}  \hat{C}_2\wedge \hat{B}_2
&=&
-\frac14 \int {\rm d}^4 x\,  \sqrt{-g}\, \epsilon^{\mu\nu\rho\sigma}C_{AB}\frac{\partial Y_{\rm (b)}^A}{\partial x^\mu}\frac{\partial Y_{\rm (b)}^B}{\partial x^\nu}B_{CD}\frac{\partial Y_{\rm (b)}^C}{\partial x^\rho}\frac{\partial Y_{\rm (b)}^D}{\partial x^\sigma} 
\nonumber
\\
& =
& -\frac{1}{4T_3^2} \int {\rm d}^4 x\, \sqrt{-g}\,  C_{IJ} B_{KL}\, \epsilon^{\mu\nu\rho\sigma}\partial_\mu\phi^I
\partial_\nu \phi^J\partial_\rho\phi^K\partial_\sigma \phi^L
\label{C2a}
\ea
where we have used the brane embedding given in Eq.~(\ref{brane-embed}) as well as the fact that $\hat{C}_2$ has only $\c$ indices.
The next term is proportional to
\beq
\int_{\rm brane}  \hat{C}_2\wedge F_2
=-\frac{1}{4T_3} \int {\rm d}^4 x \, \sqrt{-g}\, C_{IJ}\, \epsilon^{\mu\nu\rho\sigma}\partial_\mu \phi^I
\partial_\nu \phi^J F_{\rho\sigma} \, .
\label{C2b}
\eeq
We do not write explicitly the two following terms involving $\hat{B}_2\wedge \hat{B}_2$ and $\hat{B}_2\wedge F_2$, since they are analogous to (\ref{C2a}) and (\ref{C2b}), respectively.
The final term, involving $C_0$, is given by
\beq
\int_{\rm brane}  C_0 \, F_2\wedge F_2=
 -\frac14\int {\rm d}^4 x\,  \sqrt{-g}\, C_0 \epsilon^{\mu\nu\rho\sigma}F_{\mu\nu}F_{\rho\sigma}=-\frac12\int {\rm d}^4 x \, \sqrt{-g}\, C_0 F_{\mu\nu}\tilde{F}^{\mu\nu}\, ,
\label{C0a}
\eeq
where we have introduced the dual of the field strength $\tilde{F}^{\mu\nu}=\half  \epsilon^{\mu\nu\rho\sigma}F_{\rho\sigma}$. 
As is clear from the last equality, we refer to $C_0$ as the axion.

\subsection{Full action}

In the rest of this paper, we thus work with the four-dimensional effective action given by 
\ba
S &=&  \int {\rm d}^4 x \sqrt{-g} \,{}^{(4)}{{ R}} + S_{\rm brane}
\nonumber
\\
&=& 
 \int {\rm d}^4 x \sqrt{-g}\left[ {}^{(4)}{{ R}}   -\frac{1}{f(\bfphi^I)}( \sqrt{{\cal D}} -1 )  - V \right] + 
 S_{\rm WZ}^{[2]},
 \label{action}
\ea
where we have set  $M_{{\rm P}} \equiv \left(8\pi G\right)^{-1/2}=1$,  ${\cal D}$ is given explicitly in Eq.~(\ref{D-explicit}), and we have defined the potential
\be
V \equiv T_3 {\cal V} + \frac{1}{f}
\ee
so that in the small velocity limit for the scalar field, the action reduces to the usual difference between the kinetic energy and potential energy. In the following we will keep $V$ general so that it can incorporate other terms coming, for example, from moduli stabilization effects or interactions with other branes.

\section{Background evolution}
\label{sec:background}

In this section, we assume the spacetime to be  homogeneous and isotropic, and described by a spatially flat  FLRW (Friedmann-Lema\^itre-Robertson-Walker) geometry with metric
\be
g_{\mu \nu}\, dx^{\mu} dx^{\nu} = -dt^2 + a(t)^2 d{\bf x}^2\, ,
\label{bck}
\ee
where $t$ is cosmic time.

Since there is no prefered spatial direction, the scalar fields can be only time-dependent, $\p^I = \pb^I(t)$, and the field strength on the brane must vanish, $F_{\mu \nu}=0$.
This implies that the tensor $\C^{\mu}_{\,\,\nu}$, defined in Eq.~(\ref{def-A}), vanishes (since $b_{IJ} \dot \pb^I \dot \pb^J   =0$ by antisymmetry of $b_{IJ}$) and that the only non-zero component of the 
matrix $\S$ is $S_{\; 0}^0$. As a result, there are only two non-zero terms in the determinant ${\cal D}$ of Eq.~(\ref{D}), and they are the first two terms of ${\cal D}_S$.
Thus the DBI action in (\ref{action}) reduces to the simple form 
\beq
S_{\rm DBI}= -\int {\rm d}^4 x \, \sqrt{-g} \frac{1}{f(\pb^I)}\sqrt{1- f\, G_{IJ} \dot \pb^I \dot \pb^J}\,
\eeq
where a dot denotes a derivative with respect to $t$. 
Consider now the WZ terms given in Eq.~(\ref{WZexpanded}). 
All  terms involving $F_{\mu \nu}$ are automatically zero. Morever, the terms with 
$\hat{C}_2\wedge \hat{B}_2$ and  $\hat{B}_2\wedge \hat{B}_2$ also vanish, by antisymmetry (see Eq.~(\ref{C2a})). Hence $S_{\rm WZ}^{[2]}$ vanishes on the background.  
Thus, to summarize, the background dynamics is described by the four-dimensional action
\beq
S_0 =  \int {\rm d}^4 x \sqrt{-g}\left[ {}^{(4)}{{ R}}   -\frac{1}{f(\pb^I)} \left( \sqrt{1- f\, G_{IJ} \dot \pb^I \dot \pb^J} - 1 \right)   -V(\pb^I) \right]
\eeq
 which is independent of the bulk form fields.
Let us now define 
\ba
\dot{\sigma} &\equiv& \sqrt{G_{IJ} \dot \pb^I \dot \pb^J }
\label{def-sigmadot}
\ea
 (note that $\dot \s$ is not the time derivative of any field in general but merely a notational convenience)
as well as
\be
c_s \equiv \sqrt{1- f\dot{\sigma}^2},
\ee
which will later be interpreted as  the propagation speed of the perturbations, i.e.~as an effective sound speed. Then the background equations of motion are
\beq
H^2 = \frac{1}{3} \left(V+ \frac{1-c_s}{f c_s} \right)
\eeq
\beq
\dot{H} = - \frac{\dot{\sigma}^2}{2c_s}
\eeq
\beq
 \ddot \pb ^I+\Gamma^I_{JK}\dot \pb^J \dot \pb^K +\left(3H-\frac{\dot
c_s}{c_s}\right)\dot \pb^I + c_s G^{IJ}
\left(V_{,J}-\frac{f_{,J}}{f^2}\frac{\left(c_s-1 \right)^2}{2 c_s}
\right)=0\,
\eeq
 where $\Gamma^{I}_{JK} $ is the Christoffel symbol constructed from the field space metric $G_{IJ}$.

\section{Dynamics of linear perturbations}
\label{sec:secondorder}

We now study linear perturbations.  Their dynamics is described by the action quadratic in the perturbations, obtained by expanding 
Eq.~(\ref{action}) to second order about the homogeneous and isotropic background described in the previous section.

In addition to the scalar fields describing the brane position in the $\c$ dimensions, our dynamical system 
contains general relativity as well as  a non-linear $U(1)$ gauge theory. As a consequence, the linear dynamics is described by the perturbations of the scalar fields $\p^I$, of the metric $g_{\mu \nu}$ and of the gauge field $A_{\mu}$, which are all coupled. 
The perturbations can be decomposed into scalar, vector and tensor modes, following the standard terminology
in cosmological perturbation theory. These three sectors are completely decoupled  at the linearized level and 
 we will not consider tensor perturbations, i.e.~gravitational waves, as they obey the usual equations. 

It is  convenient to work in the Arnowitt-Deser-Misner (ADM) formalism, in which the metric is written in the form
\beq
ds^2=-N^2 dt^2 +h_{ij} (dx^i+N^i dt)(dx^j+N^j dt)\, .
\label{metric}
\eeq
The gauge invariance of general relativity under spacetime diffeomorphisms manifests itself in the fact that the lapse function $N$  and the shift vector $N^i$ are not dynamical degrees of freedom (they appear without time derivatives  in the action). The variation of this action with respect to the lapse and shift, yields, respectively, the energy constraint and the momentum constraint.
Similarly, we write the four-dimensional vector potential $A_{\mu}$ in `3+1' form 
\beq
A_{\mu} \equiv  \left( \A, \Ai  \right)\, 
\eeq
where, since $A_{\mu}$ vanishes on the background (up to gauge transformations), $\A$ and $\Ai$ 
are perturbations.
On imposing the Coulomb gauge $\partial^i A_i=0$, it follows that $A_i$ is transverse and contains only two vector-like modes. The scalar time component $\A$ is not dynamical and the variation of the action with respect to it leads to a third constraint. 

In the following, we work in the flat gauge 
\beq
h_{ij}=a^2(t) \delta_{ij} \, ,
\eeq 
where we denote the scalar field perturbations as $Q^I$:
\beq
\p^I = \pb^I + Q^I \, .
\eeq
We also decompose the lapse and shift into scalar and vector perturbations 
\beq
N = 1 + \delta N, \qquad  N_i = \psi_{,i}+\bar{N}_i, \qquad  \partial^i \bar{N}_i=0\,.
\eeq

We now expand the full action up to second order in the scalar perturbations $\delta N, \psi, \A$ and $Q^I$ and vector perturbations $\bar{N}_i, \Ai$. 
On using the scalar components of the energy, momentum and gauge field constraints, we can express $\delta N$, $\psi$ and $\A$ in terms of the  scalar field perturbations $Q^I$, which 
represent the true scalar degrees of freedom. Similarly, the vector-like part of the shift $\bar{N}_i$ will be 
determined in terms of $\Ai$, via the momentum constraint equations, so that the true vector degrees of freedom will 
be the  two polarizations of the brane gauge field.

\subsection{DBI determinant}

As an intermediate step in the calculation of the second-order action, it is necessary to expand, up to second order, the determinant which appears in the DBI Lagrangian.  One can either use Eq.~(\ref{D-explicit}) or more directly work with the DBI determinant of the form given in Eq.~(\ref{L}):
\ba
S_{\rm DBI} &=& -T_3\int {\rm d}^4 x\,  e^{-\d}\sqrt{-\det{ \left({\hat \gamma}_{\mu \nu}+\Bc_{\mu \nu}\right) }} \nn
\\
&=& -T_3\int {\rm d}^4 x\,  e^{-\d}\sqrt{-\hat \gamma}   \sqrt{ \det{(\delta^{\mu}_{\,\, \nu}+ \hat \gamma^{\mu \lambda} \Bc_{\lambda \nu})}} \, .
\ea
Expanding to second order yields 
\bea
S_{{\rm DBI}\,(2)}&=&-T_3\int {\rm d}^4 x\,\left(  \delta^{(2)} \left(  e^{-\d}\sqrt{-\hat \gamma}  \right)+ e^{-\bar \d}\sqrt{-\gamma} 
\frac{1}{4}\Bc_{\mu \nu}\gamma^{\mu \alpha} \gamma^{\nu \beta}\Bc_{\alpha \beta} \right)
\label{SDBI2}
\eea
where $\gamma_{\mu \nu}$ is the background induced metric
\beq
{\gamma}_{\mu \nu} {\rm d}x^\mu {\rm d}x^\nu=\frac{ e^{\Phi/2}}{\sqrt{fT_3}} \left(-c_s^2 {\rm d} t^2+a^2(t) {\rm d}\vec{x}^2\right) \, .
\label{induced}
\eeq
  (From now on we drop bars on background quantities when there is no ambiguity). 
The perturbations of $\gamma_{\mu \nu}$ are not relevant here since $\Bc$ is already first order. We have already computed the first term of Eq.~(\ref{SDBI2}) in \cite{Langlois:2008qf} and thus concentrate on the second term
which, modulo the coupling to the dilaton, is very similar to the usual electromagnetic action, though now the generalized field strength 
$\Bc_{\mu \nu}$ couples to the induced metric $\gamma_{\mu\nu}$ rather than $g_{\mu\nu}$ (this can be seen as a specific manifestation of the so-called open string metric \cite{openstringmetric}).

The non vanishing components of $\Bc$, at linear order, are given by
\bea
{}^{(1)}\Bc_{0i}&=&2\pi \alpha' \left[\frac{1}{\fF} \KR_{IJ}\dot\p^I\partial_iQ^J+F_{0i} \right] = 2\pi \alpha' \left[-\frac{1}{\fF} \partial_i\Lambda + \dot A_i \right]
\nonumber
\\
{}^{(1)}\Bc_{i j}&=&2\pi \alpha' F_{i j}=2\pi \alpha' \left(\partial_i A_j-\partial_j A_i\right),
\label{eq:first-order-F}
\eea
where all scalar modes appear in the combination
\beq
\Lambda\equiv -\KR_{IJ}\dot\p^I Q^J+ \fF A_{0}.
\eeq
On substituting the 
induced metric (\ref{induced}) and the components (\ref{eq:first-order-F}) into the second-order action (\ref{SDBI2}), 
one finds that the scalar contribution is given by
\be
S_{(2){\rm scalar}}^{\rm DBI} \supseteq \frac{1}{2}  \int {\rm d}t \,\dn{3}{x}\, \frac{a}{f c_s} (\partial\Lambda)^2. 
\label{ssscalar}
\ee
and the vector contribution by
\begin{eqnarray}
S_{(2){\rm vector}}^{\rm DBI}&=& \frac{T_3(2 \pi \alpha')^2}{2}  \int {\rm d}t \,\dn{3}{x}\,  \left[   e^{-\d}\frac{a^3}{c_s} \left( \frac{1}{a^2}\dot{A_i}\dot{A_i}+\frac{c_s^2}{a^4}A_ i\Delta A_i \right) \right] \, .
\label{2d-order-action-A}
\end{eqnarray}
Here and in the following, we use the convention that repeated (lower) spatial indices are summed over: for example, $A_ i\Delta A_i \equiv \sum_{i} A_ i\Delta A_i$.

\subsection{WZ terms}
\label{subec:2}

We now examine the contribution of the WZ terms to the second-order action.  To lowest order in perturbations, the term $\int \hat{C}_2\wedge \hat{B}_2$ given in Eq.~(\ref{C2a}) is third order (since the partial derivatives $\partial_\mu \phi^I$ are first-order except for $\mu=0$). Hence this term does not contribute to the second order action. The same conclusion applies to the term $\int C_0 \hat{B}_2\wedge \hat{B}_2$.
The term $\int \hat{C}_2\wedge F_2$ given in Eq.~(\ref{C2b}) yields, to second-order, 
\beq
\int \hat{C}_2\wedge F_2 \propto \int d^4x \, a^3  C_{IJ}\dot\p^I \epsilon^{ijk}\partial_iQ^J F_{jk}= \int d^4x\, \partial_i\left( a^3 C_{IJ}\dot\p^I \epsilon^{ijk}
Q^J F_{jk}\right) = 0
\eeq
up to total spatial derivative (our sign convention is 
$\epsilon^{ijk}=-\, \epsilon^{0ijk}$). A similar result holds for the term $\int C_0 \hat{B}_2\wedge F_2$.
Finally, we must consider (see Eq.~(\ref{C0a}))
\bea
\int C_0 F_2\wedge F_2 &=&
-\frac12\int {\rm d}^4 x \sqrt{-g}\,  C_0 F_{\mu\nu}\tilde{F}^{\mu\nu}
= 
-\frac12\int {\rm d}^4 x \sqrt{-g} \, C_0  \epsilon^{\mu\nu\rho\sigma} \partial_\mu \left({F}_{\nu\rho}A_\sigma \right)
\nn
\\
&\approx&  \int {\rm d}^4 x \, a^3 \dot{C}_0 \epsilon^{ijk}  (\partial_i A_j) A_k 
 \label{ass}
 \eea
where the last equality is obtained by integrating by parts and restricting to second order. 
In conclusion, the contribution to the second-order action coming from the WZ terms contains only {\it vector} perturbations and reads
\begin{eqnarray}
S_{(2){\rm vector}}^{\rm WZ}&=& -\frac{T_3(2 \pi \alpha')^2}{2}  \int {\rm d}t \,\dn{3}{x}\, a^3 \dot{C}_0 \, \epsilon^{ijk}  (\partial_i A_j) A_k   \, .
\label{vector-WZ}
\end{eqnarray}

\section{Linear scalar perturbations}
\label{sec:linear-scalar}

As discussed above, the new terms in the second order {\it scalar} action come only from the DBI determinant and depend on $A_0$ through $(\partial\Lambda)^2$.  Thus, if we now vary the action with respect to $A_0$, we simply obtain the constraint $\partial^2\Lambda=0$ or
\beq
\Lambda=0 \qquad \Longleftrightarrow \qquad A_0 = \frac{1}{\fF} \KR_{IJ}\dot\p^I Q^J.
\label{eq:A}
\eeq
(Note that $A_0$ vanishes in the case of a single field but is non-zero in general.)  Thus the contribution scalar second-order action Eq.~(\ref{ssscalar}) vanishes.  

As a consequence, the energy and momentum constraints are exactly the same as those of  \cite{Langlois:2008qf}, and  the linear expressions for $\delta N$ and $\psi$ are
\ba
\delta N &= & \frac{1}{2 H c_s} \dot \p_I Q^I \, ,
\label{alpha}
\\
\psi &=& -\frac{a^2}{2H} \partial^{-2} \left[\frac{1}{c_s^3} \dot \p_ I\left( \dot Q^I+\Gamma^I_{JK}\dot \p^J Q^K \right)+V_{,I}Q^I+\frac{(1-c_s)(1+c_s-2 c_s^2)}{2 f^2 c_s^3}f_{,I} Q^I  \right.
 \cr &&
 + \left. \left(3 H^2 -\frac{\dot{\sigma}^2}{2c_s^3} \right)\frac{\dot \p_ I Q^I}{H c_s} \right] \, ,
\label{beta}
\ea
(as well as $\bar N_i = 0$).
Hence on substituting Eq.~(\ref{eq:A}) as well as Eqs.~(\ref{alpha})-(\ref{beta}) back in the scalar second-order action, all new terms arising from the bulk forms  vanish, and one obtains exactly the same second-order action as in our previous works \cite{Langlois:2008qf,Langlois:2008wt}. 

Let us summarize those results, which will be useful in Sec.~\ref{sec:GW-constraints}. First define the unit vector (with respect to the metric $G_{IJ}$) tangent to the background trajectory in field space
\beq
e^I \equiv \frac{\dot \p^I}{\dot{\sigma}}\, 
\eeq
where $\dot{\sigma}$ was defined in Eq.~(\ref{def-sigmadot}),
and 
\beq
\tilde{G}_{IJ}=\frac{1}{c_s^2}e_I e_J +\left(G_{IJ}-e_I e_J \right)\,.
\label{tilde-G}
\eeq
 The second-order action for the scalar perturbations $Q^I$ then takes the form \cite{Langlois:2008qf}
\begin{eqnarray}
S_{(2){\rm scalar}}&=& \half \int {\rm d}t \,\dn{3}{x}\,    a^3 \left[ \frac{1}{c_s} \left(\tilde{G}_{IJ}
 \mathcal{D}_t Q^I \mathcal{D}_t Q^J - c_s^2 \tilde{G}_{IJ}h^{ij} \partial_i Q^I \partial_j Q^J \right)
  - {\tilde{\cal M}}_{IJ}Q^I Q^J  \right.
 \cr &&
 + \left.  \frac{f_{,J} \dot{\sigma}^2}{c_s^{3}}\dot \p_I Q^J \mathcal{D}_t Q^I  \right]\,,
\label{2d-order-action}
\end{eqnarray}
where we have introduced the time covariant derivative $\mathcal{D}_t Q^{I} \equiv \dot{Q}^I + \Gamma^{I}_{JK} \dot{\phi}^J Q^{K}$ (and $\mathcal{R}_{IKLJ}$ will denote the Riemann tensor associated to $G_{IJ}$).
The mass matrix which appears above is
\bea
{\tilde{\cal M}}_{IJ} =  \mathcal{D}_I \mathcal{D}_J V-\frac{(1-c_s)^2}{2 c_s }\frac{ \mathcal{D}_I \mathcal{D}_J f}{f^2}-\frac{(1-c_s)^3(1+3 c_s)}{4 c_s^3}\frac{f_{,I}\, f_{,J}}{f^3}
+2 \dot{H} \mathcal{R}_{IKLJ}e^K e^L  \nonumber \\
+ \frac{(1-c_s^2)^2}{2 c_s^4 f^2 H}f_{,(I} \dot{\p}_{J)}+ \frac{\dot{H}}{2H^2c_s^4}\left(1-c_s^2\right)\dot \p_I \dot \p_J
-\frac{1}{a^3}\mathcal{D}_t\left[\frac{a^3}{2H c_s^4}\left(1+c_s^2 \right)\dot \p_
I \dot \p_J\right] .
\label{Interaction matrix}
\eea
The fact that the time and spatial gradient terms in Eq.~(\ref{2d-order-action}) are multiplied by the same factor $\tilde{G}_{IJ}$ implies that \textit{all} scalar perturbations propagate at the same speed, namely the speed of sound $c_s$. 

We can gain a better intuition for the dynamics of perturbations described by the action (\ref{2d-order-action}) by restricting our attention to a two-field system, $I=1,2$.
Then one can unambiguously decompose perturbations into (instantaneous) 
adiabatic and entropic modes by projecting respectively, parallel and perpendicular to the background trajectory in field space. In other words, we introduce the basis $\{e_\sigma, e_s\}$ where 
$e^I_\sigma=e^I$, and $e^I_s$ is the entropy unit vector orthogonal to $e^I_\sigma$:
\beq
e_\sigma^I\equiv e^I, \qquad G_{IJ}e_s^I e_s^J=1, \qquad G_{IJ}e_s^I e_\sigma^J=0.
\eeq
It is then convenient, after going to conformal time $\tau = \int {{\rm d}t}/{a(t)}$, to work in terms of the canonically normalized fields given by
\be
v_{\s}=\frac{a}{c_s^{3/2}} \, e_{\s I}Q^I \,,\qquad \,v_{s}=\frac{a}{\sqrt{c_s}}\, e_{s I} Q^I\,.
\label{v}
\ee
In particular, in terms of these the scalar action Eq.~(\ref{2d-order-action}) simplifies remarkably \cite{Langlois:2008mn,Langlois:2008qf} to
\bea
\label{S_v}
S_{(2){\rm scalar}}&=&\frac{1}{2}\int {\rm d}\tau\,  {\rm d}^3x \left[ 
  v_\s^{\prime\, 2}+ v_s^{\prime\, 2} -2\xi v_\s^\prime v_s-c_s^2 \left[(\partial v_\s)^2 + (\partial v_s)^2 \right] 
+\frac{z''}{z} v_\s^2  \right.
 \cr &&
 + \left.  \left(\frac{z_s''}{z_s}-a^2 \mu_s^2\right) v_s^2+2\, \frac{z'}{z}\xi v_\s v_s \right]
\eea
where a prime denotes a derivative with respect to conformal time, leading to the equations of motion (in Fourier space):
\begin{eqnarray}
v_{\s}''-\xi v_{s}'+\left(c_s^2 k^2-\frac{z''}{z}\right) v_{\s} -\frac{(z \xi)'}{z}v_{s}&=&0\,,
\label{eq_v_sigma}
\\
v_{s}''+\xi  v_{\s}'+\left(c_s^2 k^2- \frac{z_s''}{z_s}+a^2\mu_s^2\right) v_{s} - \frac{z'}{z} \xi v_{\s}&=&0\,.
\label{eq_v_s}
\end{eqnarray}
Here we have introduced the two background-dependent  functions 
\beq
z=\frac{a \dot \s }{H c_s^{3/2}}, \qquad z_s=\frac{a}{\sqrt{c_s}} \, ;
\eeq
the coupling between $v_\s$ and $v_s$ depends on 
\beq
\xi = -a \sqrt{\frac{f}{1-c_s^2}}\left[\frac{(1-c_s)^2}{f^2}f_{,s}+(1+c_s^2)V_{,s}\right]
\label{11}
\eeq
(where $V_{,s} \equiv e_s^I V_{,I} \,,V_{;ss} \equiv e_s^I e_s^J {\cal D}_I {\cal D}_J V$ and similarly in the following), and finally the effective mass appearing above is given by
\bea
\mu_s^2 &\equiv& c_s V_{;ss}-\frac{f}{1-c_s^2}V_{,s}^2-\frac{(1-c_s)^3}{4(1+c_s)f^3} f_{,s}^2-\frac{(2+c_s)(1-c_s)}{(1+c_s)f}f_{,s} V_{,s} \nonumber \\
 && - \frac{(1-c_s)^2}{2f^2} f_{;ss}+\half \dot \s^2{{\cal R}}_G\,.
\label{mus2}
\eea
(${{\cal R}}_G$ is the scalar Riemann curvature in field space.) 

Following the standard procedure 
(see e.g.~\cite{mfb,Langlois:2004de}),  
Eqs.~(\ref{eq_v_sigma}) and (\ref{eq_v_s}) can be used as the starting point to quantize the perturbations and derive the scalar and tensor spectra generated during two-field DBI inflation. Note that the amplification of the 
quantum fluctuations occurs when the scales cross out the {\it sound horizon}, i.e.~when $kc_s= aH$, as in $k$-inflation \cite{Garriga:1999vw}. 
The scalar spectrum depends on the  slow-varying parameters
\beq
\epsilon \equiv -\frac{\dot H}{H^2}, \qquad \eta=\frac{\dot \epsilon}{H \epsilon}\,, \qquad s=\frac{\dot c_s}{H c_s}\,,
\label{eq:slow-roll}
\eeq
which are assumed to be 
small. 
The scalar amplitude, expressed in terms of the comoving curvature perturbation ${\cal R}= v_{\s} / z$ is given by
\beq
{\cal P}_{\cal R}=  \frac{{\cal P}_{\cal R_{*}}}{{\cos^2} \Theta} =\left(\frac{H^2}{8\pi^2 \epsilon \, c_s}\right)_* \frac{1}{{\cos^2} \Theta},
\label{strike}
\eeq
where
the index $*$ means that the corresponding quantity is evaluated  
  at sound horizon crossing, 
  and the parameter
$\Theta$ quantifies the amplification of the curvature perturbation {\it after} horizon crossing as a consequence of the transfer of entropy perturbations into curvature perturbations. This feeding of the curvature perturbation by the entropy modes is a characteristic feature of multi-field inflation, either with standard kinetic terms or 
with non standard kinetic terms (see \cite{sy} and \cite{Lalak:2007vi,Vincent:2008ds,Ashoorioon:2008qr,Chen:2008ada,Tye:2008ef} for other recent illustrations in the context of string inflation). 
For $\Theta=0$ there is no transfer and Eq.~(\ref{strike}) reduces to the standard result of single field DBI inflation. From Eq.~(\ref{strike}) the spectral index is given by \cite{Langlois:2008qf}
\beq
n_{\cal R}-1\equiv \frac{{\rm d\,ln}{\cal P_{{\cal R}}}}{{\rm d\, ln\,}k}=
-2\epsilon_*- \eta_* -s_*-\alpha_{*}{\rm sin}(2 \Theta)-2\beta_*{\rm sin^2} \Theta
\label{index}
\, 
\eeq
with
\be
\alpha= \frac{\xi}{a H} \; ,\qquad \beta \simeq \frac{s}{2}-\frac{\eta}{2}-\frac{1}{3H^2}\left(\mu_s^2+\frac{\Xi^2}{c_s^2}\right), \quad \Xi\equiv\frac{c_s}{a}\xi,
\label{coefficients}
\ee 
and where we have  kept only the leading order terms in the expression for $\beta$.
Finally, the power spectrum for tensor modes in multifield DBI inflation is unmodified relative to standard multifield inflation, and hence 
the tensor to scalar ratio is given by
\beq
r \equiv \frac{{\cal P}_{\cal T}}{{\cal P}_{\cal R}}=16 \, \epsilon \, c_s \,{\cos^2} \Theta.
\label{r}
\eeq
We will return to the above quantities in Section \ref{sec:GW-constraints}, where we discuss various theoretical and observational
constraints.

\section{Linear vector perturbations}
\label{sec:vector}

In this section we consider the vector perturbations $\Ai$ which,  
 at linear order, are decoupled from scalar perturbations. 
 
The second order action for vector perturbations has two parts: that coming from the DBI determinant given in Eq.~(\ref{2d-order-action-A}) as well as the contribution from the WZ term given in Eq.~(\ref{vector-WZ}).  
In the same way as for the scalar perturbations in the previous section, we work in conformal time $\tau$ and introduce the
 canonically normalized fields
\be
v_i=(2 \pi \alpha') \sqrt{T_3} \left( \frac{e^{- \d/2}}{\sqrt{c_s}} \right) A_i \, .
\label{vcan}
\ee
After an integration by parts, the second-order action for vector perturbations becomes
\begin{eqnarray}
S_{(2){\rm vector}}= \half \int {\rm d}\tau \,\dn{3}{x} \left[v_i'^2  - c_s^2 (\partial v_ i)^2
+\frac{\chi''}{\chi}  v_i^2 
-\frac{1}{\chi^2}C'_0 \epsilon^{ijk}v_i\partial_j v_k
 \right]\,
 \label{snow}
\end{eqnarray}
where 
\beq
\chi\equiv \frac{e^{-\Phi/2}}{\sqrt{c_s}} \, .
\eeq
(Action (\ref{snow}) for vector perturbations if the analogue of action~(\ref{S_v}) for scalar perturbations.) When $C'_0 = 0$, each Fourier mode satisfies the equation of motion
\beq
v''_i +\left(k^2 c_s^2-  \frac{\chi''}{\chi}    \right) v_i= 0\, .
\label{hmmm}
\eeq
 One can see that the vector perturbations again propagate at the sound speed $c_s$.  Eq.~(\ref{hmmm}) is very similar to those for the scalar perturbations Eqs.~(\ref{eq_v_sigma}) and (\ref{eq_v_s}) --- however, while scalar perturbations are amplified, there is no amplification of the vector modes. The reason is that the term in $\chi''/\chi$ can simply be absorbed by a suitable redefinition of the time derivative, as was already noticed in \cite{Bamba:2008my}.
A more intuitive way to understand this result is to 
rewrite the homogeneous induced metric given in Eq.~(\ref{induced}) in a form which is manifestly conformal to the Minkowski metric
 \beq
{\gamma}_{\mu \nu} {\rm d}x^\mu {\rm d}x^\nu=
\frac{ e^{\Phi/2}}{\sqrt{fT_3}} a^2\left(-{\rm d}\eta^2+{\rm d}\vec{x}^2\right)\, ,
\eeq
and which naturally defines a new time variable 
 $\eta$ by $c_s {\rm d}t=a {\rm d} \eta$.
In terms of $\eta$ and the canonically normalized field
\beq
\psi_i =
\sqrt{c_s} v_i = 2 \pi \alpha' \sqrt{T_3} e^{-\half  \d} A_i
\eeq
the action indeed simplifies to
\begin{eqnarray}
S_{(2)\, {\rm vector}}= \half \int {\rm d}\eta \,\dn{3}{x}\,  
\left(\frac{{\rm d} \psi_i}{{\rm d} \eta}\frac{{\rm d} \psi_i}{{\rm d} \eta}+\psi_ i\Delta \psi_i
+
e^{{\d}/{2}}  \frac{{\rm d}^2 e^{-{ \d}/{2}}  }{{\rm d} \eta^2} \psi_i \psi_i 
-
\frac{{\rm d} C_0 }{{\rm d} \eta} e^{\d} \epsilon^{ijk} \psi_i \partial_j \psi_k \right) \, .
\label{psi_i_equation}
\end{eqnarray}
As usual we now expand the vector field in terms of annihilation and creation operators, $ {\hat a}_{p \boldsymbol{k}}$ and ${\hat a}^{\dagger}_{p \boldsymbol{k}} $
\bea
\psi_i(\eta,x^i)&=& \int \frac{ {\rm d}^3k}{(2\pi)^{3/2}} \sum_{p=1}^{2} \epsilon_{i p}(\boldsymbol{k}) \left[ \psi_{p}(\eta) {\hat a}_{p \boldsymbol{k}} e^{i \boldsymbol{k} \cdot \boldsymbol{x}}   +  \psi_{p}^{*}(\eta) {\hat a}^{\dagger}_{p \boldsymbol{k}} e^{-i \boldsymbol{k} \cdot \boldsymbol{x}}        \right],
\label{Fourier-expansion-vectot}
\eea
where the transverse polarization vectors $\epsilon_{i p}(\boldsymbol{k})$ defined by
\bea
\sum_{p=1}^{2} \epsilon^{i}_{p}(\boldsymbol{k}) \epsilon_{j p}(\boldsymbol{k})=\delta^i_j-\delta_{jl} \frac{k^i k^l}{k^2}
\label{polarization}
\eea
are introduced for a consistent quantization in the Coulomb gauge.
The last term in Eq.~(\ref{psi_i_equation}) introduces a mixing between the two linear polarisation degrees of freedom of the photon $\psi_{p}$. This interaction can be diagonalized by considering instead the two circular polarisation degrees of freedom
\bea
\psi_{(\pm)} &=& \psi_1 \pm i  \psi_2\,
\eea
so that the corresponding equations of motion are 
\beq
\frac{{\rm d}^2 \psi_{(\pm)}}{{\rm d} \eta^2}+\left(k^2- e^{\half \d} \frac{{\rm d}^2 e^{-\half  \d}}{{\rm d} \eta^2}\mp k \frac{{\rm d} C_0 }{{\rm d} \eta} e^{\d} \right)\psi_{(\pm)}=0\,.
\label{eom-circular}
\eeq
Thus, when 
 the dilaton and axion are $\eta$-independent, or equivalently homogeneous in the compact dimensions, there is no amplification of the vector field.  The underlying reason is that the action, written in terms of the induced metric, is still conformally invariant (see Eq.~(\ref{SDBI2})) in the same way that standard electromagnetic theory is conformally invariant \cite{Parker:1968mv}.  
 From Eq.~(\ref{eom-circular}), on the other hand, a time-varying dilaton and/or axion causes the vector field to be amplified.

Indeed, other than the replacement of $\eta$ by conformal time $\tau$, Eq.~(\ref{eom-circular}) is identical to that obtained in standard electromagnetism coupled to a dilaton and an axion \cite{Ratra:1991bn,Garretson:1992vt}. In that case, Eq.~(\ref{eom-circular}) has been used to compute the amplitude of quantum fluctuations generated during inflation, with the aim of addressing the question of the generation of primordial magnetic fields (see e.g.~Refs.~\cite{Kronberg:1993vk,Grasso:2000wj,Widrow:2002ud} for reviews).   
In the context of DBI inflation,
the gauge field $A_\mu$ is localized on the D-brane responsible for inflation, and is not obviously related to our standard electromagnetic field: hence the applicability of Eq.~(\ref{eom-circular}) to magnetogenesis is not immediate.  In particular the brane responsible for inflation is not usually the standard model brane, and understanding  how perturbations of the gauge field in one throat might couple to perturbations of electromagnetic fields on the standard model brane in a (possibly) different throat, is a challenging open question.

\section{Non-Gaussianities}
\label{sec:NG}

In this section, we analyse the effects of the NS-NS and R-R bulk fields on non-Gaussianities. We thus calculate the third-order action for the scalar-type perturbations. In order to do so, it is sufficient to calculate $N$, $N_i$ and $\A$ to first order:  $\A$ is given in Eq.~(\ref{eq:A}) and $\delta N$ and $\psi$ are given respectively in Eqs.~(\ref{alpha}) and (\ref{beta}) (and $\bar N_i=0$). 

First recall that since $\Lambda=0$, the scalar part of $\Bc$ vanishes to linear order (see Eq.~(\ref{eq:first-order-F})). Furthermore since
$\Bc$ enters the DBI action  Eq.~(\ref{D-explicit}) at least quadratically, the third-order DBI action for the purely scalar perturbations $Q^I$ is exactly the same as in \cite{Langlois:2008qf,Langlois:2008wt}.  Let us now turn to the WZ action. The potential term Eq.~(\ref{pot}) and its non-linearities were already considered in our previous work, hence we focus on the remaining part, $S_{\rm WZ}^{[2]}$ given in Eq.~(\ref{WZexpanded}) and which can be rewritten as
\bea
 S_{\rm WZ}^{[2]}&=&-T_3 \left[ \int_{\rm brane} \hat{C}_2\wedge \left(\hat{B}_2 +2 \pi \alpha' F_2 \right)  \right.
 \cr &&
 + \left. \frac12 \int_{\rm brane} C_0 \left(\hat{B}_2 +2 \pi \alpha' F_2 \right) \wedge \left(\hat{B}_2 +2 \pi \alpha' F_2 \right)  \right].
\label{WZ-remaining}
\eea
Since the scalar part of $\Bc$ is at least second order in perturbations, the same is true for $\hat{B}_2 +2 \pi \alpha' F_2$. Therefore, the second line in Eq.~(\ref{WZ-remaining}) does not contribute to the third-order scalar action. As for the first term, its contribution is
\bea
S_{(3)\, {\rm scalar}}& \supset & -\frac{1}{2 T_3}\int {\rm d}^4 x \, a^3 \, C_{IJ} B_{KL} \dot \p_{I}  \epsilon^{ijk} \partial_i Q^J \partial_j Q^K \partial_k Q^L  \nonumber \\
& =&    -\frac{1}{2 T_3}\int {\rm d}^4 x \, \partial_i \left( a^3 C_{IJ} B_{KL} \dot \p_{I}  \epsilon^{ijk} Q^J \partial_j Q^K \partial_k Q^L \right)\,,
\eea
and hence is a total spatial derivative. Thus we obtain the remarkable result that \textit{the action, at second and third order in scalar perturbations, is not at all modified by the bulk forms.}
In the two-field case, the scalar third-order action in the small sound speed limit is given by \cite{Langlois:2008qf,Langlois:2008wt}

\ba
S_{(3)\, scalar}^{(\rm main)}&=&\int {\rm d}t\, {\rm d}^3x\,  \left\{ \frac{a^3}{2 c_s^5 \dot \s}\left[(\dot Q_{\s} )^3+c_s^2 \dot Q_{\s}  (\dot Q_{s} )^2\right]  \right.
 \cr
 && \left. 
 - \frac{a}{2 c_s^3 \dot \s}\left[ \dot Q_{\s} (\nabla  Q_{\s} )^2 -c_s^2 \dot{Q_{\s} }(\nabla Q_s)^2+2 c_s^2 \dot {Q_s}\nabla Q_{\s} \nabla Q_s)\right]
 \right\}
 \label{S3}
\ea
where
\beq
Q_{\s} \equiv e_{\s I}Q^I \,, \qquad Q_{s} \equiv e_{s I}Q^I
\eeq
are the instantaneous adiabatic and entropic perturbations respectively.

From  Eq.~(\ref{S3}) one can calculate the bispectrum of the curvature perturbation \cite{Langlois:2008qf}. It has  the same shape --- equilateral --- as in single field DBI inflation, but the corresponding non-Gaussianity parameter $f_{\rm NL}^{eq}$ is reduced by the multiple-field effects:
\beq
f_{\rm NL}^{eq}=-\frac{35}{108}\frac{1}{c_s^2} {\cos^2} \Theta \,\simeq -\frac{{\cos^2} \Theta}{3c_s^2} \, 
\qquad (c_s\ll 1)
\label{f_NL}
\eeq
where the transfer parameter $\Theta$ was introduced in Eq.~(\ref{strike}). 
Of course, at third-order in the action, there are non-trivial interactions between the scalars $Q^I$ and the vector modes $A_i$, which lead, via loop effects, to corrections to the spectrum and bispectrum, as was explored recently in a different context \cite{Seery:2008ms}. Such a study is beyond 
beyond the scope of the present paper and we leave it for future work.

A local-type contribution to primordial non-Gaussianities can also be expected, due to the nonlinear evolution of perturbations on superhorizon scales. It could be estimated by using the non-linear formalism developed in \cite{Langlois:2006vv}
and recently extended to a wide class of multifield inflationary models with non-standard kinetic terms in \cite{RenauxPetel:2008gi}.

Besides its amplitude and shape, the scale dependence of non-Gaussianities is also an interesting probe of the early universe physics and its effects on cosmological structures, in particular, have been studied recently in e.g.~\cite{LoVerde:2007ri}. 
We therefore define
\beq
n_{{\rm NG}}^{eq}=\frac{{\rm d}\, \ln\, f_{{\rm NL}}^{eq}}{{\rm d\, ln} \, k}\,,
\eeq
which is insensitive to the model-dependent non-Gaussianities of local type. From Eq.~(\ref{f_NL}), it follows that to leading order 
\beq
n_{{\rm NG}}^{eq}=-2 s_{*}+\alpha_* \sin(2 \Theta) +2 \beta_* \sin^2\Theta
\label{running}
\eeq
which reduces to the well known single field result for $\Theta=0$. We now turn to the observational constraints on multi-field DBI inflation.

\section{Gravitational wave constraints on DBI inflation}
\label{sec:GW-constraints}

The constraints on the tensor to scalar ratio $r$ in the case of {\it single-field} DBI inflation are very severe, essentially meaning that the simplest single-field UV DBI inflation models are ruled out. 
Indeed, two bounds on $r$ have been obtained \cite{Baumann:2006cd,Lidsey:2007gq}.  The first is a {\it lower bound}, valid when $c_s\ll 1$, and given by $r \, \gsim \, 0.1 (1-n_{\cal R})$. For $n_{\cal R}$ given by the best fit value obtained by WMAP5 \cite{Komatsu:2008hk} this becomes $r \, \gsim \, 10^{-3}$.  The second bound is an {\it upper bound}, typically $r \, \lsim \, 10^{-7}$, applicable in standard type IIB compactifications.  Clearly these are not easily compatible! 
More elaborate (though still effectively single-field) models have been considered in order to evade these bounds, involving wrapped branes \cite{Becker:2007ui,Kobayashi:2007hm,Silverstein:2008sg,McAllister:2008hb,Avgoustidis:2008zu} or multiple branes \cite{Krause:2007jr,Huston:2008ku} rather than a single D3-brane.  Here, however, we stick with a single D3-brane, but consider the multi-field aspects: we show that the constraints mentioned above are not incompatible in this more general context.

We now discuss the origin of the upper bound, originally due to Lyth \cite{Lyth:1996im}, in the context of multi-field DBI inflation.
The starting point is the deceleration parameter $\epsilon$ defined in Eq.~(\ref{eq:slow-roll})
 \beq
\epsilon=\frac{\dot \s^2}{2 c_s H^2 M_P^2}
\label{eps}
\eeq
where we have reinstated the Planck mass $M_P$ (previously put to unity) and 
where $\dot{\s}^2 = G_{IJ} \dot\phi^I \dot\phi^J$, as defined in Eq.~(\ref{def-sigmadot}). Note that in  
single-field DBI $\dot\sigma^2$ reduces to $\dot{\phi}^2$ where $\phi$ is the inflaton.  
Using Eq.~(\ref{r}), the tensor to scalar ratio can be written as
\beq
r=8\frac{\dot\sigma^2}{H^2 M_P^2}\cos^2\Theta=\frac{8}{M_P^2}\left(\frac{d\sigma}{dN}\right)^2\cos^2\Theta,
\label{end}
\eeq
where $N$ is the number of e-folds and ${\rm d}\sigma$ is the infinitesimal distance along the background trajectory in field space, ${\rm d}\sigma = \dot{\sigma} {\rm d}t$. Let us define 
\beq
\label{N_eff}
 N_{\rm eff}\equiv \int_0^{N_{end}} {\rm d}N\, \left(\frac{r}{r_{\rm CMB}}\right)^{1/2}\frac{1}{\cos \Theta}
 =\frac{4}{r_{\rm CMB}^{1/2}}\int_0^{N_{end}} {\rm d}N\, \sqrt{\epsilon c_s}
\eeq
where $N_{\rm end}$ in the number of e-folds from the time the present Hubble scale exits the horizon to the end 
of inflation, and  $r_{\rm CMB}$ corresponds to the observable tensor to scalar ratio. Our definition 
(\ref{N_eff}) differs from that given in \cite{Baumann:2006cd} only by the inclusion of $\cos\Theta$. Combining Eqs.~(\ref{end}) and (\ref{N_eff}), $r_{\rm CMB}$ is
thus related to the distance along the background trajectory $\Delta\s = \int_0^{N_{end}} {\rm d}\sigma$ by the expression 
 \beq
 r_{\rm CMB}=\frac{8}{N_{\rm eff}^2} \left( \frac{\Delta \s}{M_P} \right)^2. 
 \eeq 
In single-field DBI, $\Delta \sigma = \Delta \phi$, and since the field variation is limited by the size of the throat, one obtains the upper bound cited above \cite{Lidsey:2007gq}. 
With respect to single field DBI inflation, multi-field effects act in two opposite directions. The entropy-adiabatic transfer tends to increase $N_{\rm eff}^2$ (assuming the same evolution $r(N)$), which leads to a smaller $r_{\rm CMB}$. However, in warped flux compactifications with a conical throat, the field variation can now be larger than in the single-field case (corresponding to a purely radial motion)  since the size of the throat limits only  the radial displacement of the brane, not the angular displacement. 

In fact the exact size of the upper bound on $r$ is not the crucial component of this discussion. The most dramatic change comes from the second constraint (formerly a lower bound), to which we now turn.
For that purpose, note that using the definitions given in Eq.~(\ref{eq:slow-roll}), $\eta$ can be reexpressed in terms of the two other parameters $\epsilon$  and $s$, as well as the time
derivative of the warp factor $f$. Indeed, the time derivative of $\epsilon$ in Eq.~(\ref{eps}) gives
\beq
\eta=\frac{2 \ddot \s}{H \dot \s}-s+2 \epsilon= 2\epsilon -\frac{1+c_s^2}{1-c_s^2}s -\frac{\dot f}{Hf},
\eeq
where we have used the relation $\dot \s^2   = (1-c_s^2)/f$ and its derivative for the second equality.
After substitution in (\ref{index}), and using $r=16 \epsilon \cos^3\Theta/\sqrt{3|f_{\rm NL}|}$, one obtains
\bea
1-n_{\cal R}
&\simeq& \frac{\sqrt{3 |f_{{\rm NL}}| }r}{4 \cos^3\Theta}
-\frac{\dot f}{Hf} +\alpha_{*}{\rm sin}(2 \Theta)+2\beta_*{\rm sin^2} \Theta
\label{r_f}
\eea
where we have neglected a term proportional to $c_s^2s_*$,  
since $s_*$ is small and we work in the relativistic regime ($c_s^2 \ll 1$).  Note that the last two terms in (\ref{r_f}) are related to the observables $n_{{\rm NG}}^{eq}$ and $s_{*}$ through Eq.(\ref{running}).

In the UV single field case, in which $\dot{f}>0$ and $\Theta=0$, the last two terms in Eq.~(\ref{r_f}) vanish, and this leads to the lower bound
\beq
r > \frac{4}{\sqrt{3 |f_{{\rm NL}}|}}\left(1-n_{\cal R} \right)\,  \qquad ({\rm single\  field})
\label{lowerbound-single}
\eeq
From the upper bound on $|f_{{\rm NL}}|$ from WMAP5
\beq
-151 < f_{{\rm NL}}^{{\rm equil}} < 253 \qquad {\rm at}\,\,\, 95 \%\,\,\, {\rm C.L.}\,,
\eeq 
one obtains $r \, \gsim \, 0.1 \, (1-n_{{\cal R}}) \gsim 10^{-3}$.

In the multi-field case, still assuming a UV scenario with $\dot f>0$, one immediately sees that the detection of a 
deviation from scale-invariance is no longer incompatible with a very low prediction for the amount of gravitational waves.  Indeed, when 
there is a non-negligible transfer between entropy and adiabatic modes $(\Theta \neq 0)$, the last two terms of Eq.~(\ref{r_f}) can now give a significant contribution to $(1-n_{\cal R} )$, even when $r$ is too small to be detected. This shows that the very stringent constraint Eq.~(\ref{lowerbound-single}) no longer applies in a multi-field setup.

\section{Conclusions}
\label{sec:conclusion}

In the present work, we have analysed the generation of primordial
perturbations in the context of multi-field DBI inflation,
by taking into account the various  bulk form fields, which influence the
dynamics of a D3-brane moving in a  six-dimensional compact space.

For linear perturbations of the scalar type, we have shown that, in the
presence of a bulk form $B_2$, the fluctuations of the scalar fields (i.e.
fluctuations of the brane position in the compact space)  induce
fluctuations of the $U(1)$ gauge field confined on the brane. This precise
relation between the two entails an exact compensation in the second-order
scalar action between the terms coming from the perturbations of $F_2$ and
the terms coupling the scalar field perturbations to the bulk fields. As a
consequence, the second-order action expressed in terms of the true scalar
degrees of freedom is exactly the same as that obtained by ignoring the
bulk form fields and the $U(1)$ gauge field on the brane. The same
cancellation occurs in the third-order scalar action. 

A new feature is the presence of two vector degrees of freedom arising
from the $U(1)$ gauge field confined on the brane. If the dilaton and
axion are trivial, i.e. homogeneous along the compact dimensions, the
dynamics of the (linear) vector modes is governed by an action which is
quite similar to that of standard electromagnetism, with the exception
that gauge field is coupled to the induced background metric rather than
the usual FLRW metric. This property, well-known in other contexts and
which we have applied to cosmology here, explains in a transparent way why
the gauge field quantum fluctuations are not amplified, as has been
pointed out recently. Amplification is possible with a  non-trivial
dilaton or axion, as in similar  mechanisms proposed to generate
primordial magnetic fields in the very early Universe. However, since the gauge field under consideration is a priori distinct from any standard model gauge field, its relevance to magnetogenesis remains to be established.

Finally, we have extended the constraints on gravitational waves for {\it
multi-field} DBI-inflation. This analysis is
especially important as it has been shown that typical UV {\it
single-field} DBI inflation is ruled out by incompatible constraints on
the tensor to scalar ratio: on the one hand, the amount of gravitational
waves generated in most models derived from string theory is extremely
small; on the other hand, the deviation of the scalar spectrum from exact
scale invariance, which is now favoured by CMB observations, implies a
non-negligible amount of gravitational waves, at least much higher than
the  typical theoretical predictions.

For multi-field DBI inflation, the upper bound for the predicted amount of
gravitational waves can change, but no dramatically so in general.
However, the second constraint on gravitational waves is no longer valid
in multi-field inflation since the observed spectral index depends not only on $r$ 
 but also on the transfer between entropy and
adiabatic modes. Therefore, a tiny amount of gravitational waves is
perfectly compatible with a deviation from scale invariance if an
entropy-adiabatic  transfer took place while the cosmological
perturbations observed today were generated during inflation.

\label{sec:7}

\section*{Acknowledgements}
We would like to thank Emilian Dudas, Elias Kiritsis and Francesco Nitti, and particularly Costas Bachas, for useful discussions.


\end{document}